\documentclass[leqno]{llncs}

\usepackage{graphicx}
\usepackage{amsmath}

\begin{document}
\pagestyle{plain}
\pagenumbering{arabic}

\title{Epcast: Controlled Dissemination in Human-based Wireless Networks by means of Epidemic Spreading Models}
\author{Salvatore Scellato$^1$, Cecilia Mascolo$^2$, Mirco Musolesi$^3$, Vito Latora$^4$}

\institute{
Scuola Superiore di Catania\\
Via S. Nullo 5/i, 95123, Catania, Italy\\
\email{sascellato@ssc.unict.it}\\
\and
Department of Computer Science, University College London\\
Gower Street, London, WC1E 6BT, United Kingdom\\
\email{c.mascolo@cs.ucl.ac.uk}\\
\and
Department of Computer Science, Dartmouth College\\
6211 Sudikoff Laboratory, Hanover NH 03755 USA\\
\email{musolesi@cs.dartmouth.edu}\\
\and
Dipartimento di Fisica e Astronomia, Universit\`a di Catania  \\
and INFN Sezione di Catania \\
Via S. Sofia 64, 95125, Catania, Italy \\
\email{latora@ct.infn.it}
}

\maketitle
\begin{abstract}
Epidemics-inspired techniques have received huge attention in recent
years from the distributed systems and networking communities. These
algorithms and protocols rely on probabilistic message replication and
redundancy to ensure reliable communication. Moreover, they have been
successfully exploited to support group communication in distributed
systems, broadcasting, multicasting and information dissemination
in fixed and mobile networks.  However, in most of the existing work,
the probability of infection is determined heuristically, without
relying on any analytical model. This often leads to unnecessarily
high transmission overheads.

In this paper we show that models of epidemic spreading in complex
networks can be applied to the problem of tuning and controlling the
dissemination of information in wireless ad hoc networks composed of
devices carried by individuals, i.e., human-based networks. The
novelty of our idea resides in the evaluation and exploitation of the
structure of the underlying human network for the automatic tuning of
the dissemination process in order to improve the protocol
performance. We evaluate the results using synthetic mobility models
and real human contacts traces.
\end{abstract}


\section{Introduction}
\label{introduction}

Mobile human networks (i.e., ad hoc networks composed by
devices carried by individuals) can be frequently and temporarily 
disconnected. Traditional routing protocol, including the basic
flooding, fail to offer any sort of reliability when this happens.
Epidemic-style protocols instead, being store and forward approaches
and inherently delay tolerant~\cite{dtn}, allow for communication in
dynamic and mobile networks, also in presence of temporary
disconnections or network partitions.
A desired feature of the protocols is the
ability to control the information spreading.
 For example, in emergency scenarios, when the network infrastructure
has failed, it may be sufficient to send the messages only to a
percentage of the rescue team members (e.g., 50\% of the doctors).  In
other situations, there might be a need to reach all the deployed
emergency personnel with the minimum overhead to avoid to collapse the
network. Up to our knowledge, no solutions exploiting the minimal
necessary and sufficient number of replicated messages, given the
emergent network structure to guarantee a desired level of reliability
exist.


The analogy between information dissemination in mobile systems and epidemics
transmission in social systems is apparent. Information spreading can be modelled
with a simple model for disease spreading, the so-called SIR 
(Susceptible-Infected-Recovered) model~\cite{AndMay92}: 
a host is initially 
\textit{Susceptible} to new
information, then it becomes \textit{Infected} when he actually receives it, and
finally it can stop the store-and-forward dissemination process becoming
\textit{Recovered} and, therefore, immune to further infections.
Epidemics-inspired techniques have received huge attention in recent years from
the distributed systems community~\cite{EGKM04}. These algorithms and protocols
rely on probabilistic message replication and redundancy to ensure reliable
communication. Epidemic techniques were firstly exploited to guarantee consistency
in distributed databases~\cite{epidemicroutingdatabase}. More recently, these
algorithms have been applied to support group communication in distributed
systems. In particular, several protocols have been proposed for
broadcasting, multicasting and
information dissemination in fixed networks~\cite{EHGKZ03}. 

A few attempts have been made to apply epidemic based techniques for
information dissemination in mobile ad hoc
networks~\cite{epidemicrouting,CP05,BCG05}.  However, existing
epidemic algorithms do not permit to control the spreading of the
information depending on the desired reliability and the network
structure.  This is partly due to the fact that these approaches are
fundamentally based on empirical experiments and not on analytical
models: the input parameters that control the dissemination process
are selected by using experimental results and are not based on any
mathematical model. This implies that the message replication process
cannot be tuned with accuracy in a dynamic way: for instance, it is
not possible to set the parameters of the dissemination process in
order to reach only a certain desired percentage of the hosts in a
prefixed amount of time.  Moreover, these approaches do not exploit
the information on the underlying network topology
~\cite{AlBa02,BBPV05,boccaletti06}.
The use of epidemic spreading models based on the structure of the 
underlying network allows us to devise accurate mechanisms for 
controlling the message replication process. In other words, the
number of the replicas in the network and their persistence can be
tuned to achieve a desired delivery ratio.

In~\cite{MM06} we have presented initial results based on the so-called SIS
(Susceptible-Infected-Susceptible), a model of disease spreading not considering
the \textit{recovered} state.  In this paper, we propose a refined version of
the algorithm based on a SIR model. The use of SIR, in coordination with the
ability to decide to constrain the epidemics to a percentage of hosts, allows us
to lower the message overhead considerably with respect to both our previous
work and other approaches, as shown in our results section.  We present an
extended evaluation based on synthetic models and real traces of connectivity of
the Dartmouth College~\cite{dartmouth} and National University of
Singapore~\cite{singapore} campuses.



This paper is structured as follows. In Section \ref{middleware} we
describe the implementation of the middleware interface supporting the
epidemic dissemination process. Section \ref{design} presents briefly
the models of epidemic spreading in complex networks that are at the
basis of our dissemination algorithm. The implementation issues are
discussed in Section \ref{implementation}. The proposed dissemination
algorithm is evaluated analytically and by means of simulations in
Section \ref{evaluation}. Section \ref{concluding} concludes the
paper.

\section{Primitives for Controlled Epidemic Dissemination}
\label{middleware}
Our goal is to provide a primitive that allows developers to tune
information dissemination in human networks according to their specific
application requirements. Our aim is to ensure the spreading of information from
a source $A$ to a certain percentage $\Psi$ of the mobile hosts of the system
in a given interval time defined by a timeout $t^*$.

We introduce a primitive for \textit{probabilistic anycast communication} as
follows:

{ \small \begin{verbatim} epcast(message,percentageOfHosts,time) \end{verbatim}
}

\noindent where \verb+message+ is the message that has to be sent to a certain
percentage of hosts equal to the value defined in \verb+percentageOfHosts+ in a
bounded time interval equal to \verb+time+.

By using these basic primitives, more complex programming interfaces and
communication infrastructures can be designed, such as publish/subscribe systems
or service discovery protocols. 

The infectivity of the epidemics (i.e., the probability of being infected by a
host that is in the same radio range, like in human diseases spreading) can be
used to control the anycast probabilistic communication mechanism. 
Given a percentage of hosts that has to be infected equal to $\Psi$, we
are able to accurately calculate the value of the infectivity in order to obtain
an infection rate equal to a proportion of the total number of the hosts in the
network. 

As we will discuss in the next section, these primitives rely on a probabilistic
algorithm based on the transmission of a \textit{minimal}, and, at the same
time, sufficient, number of messages. 
Existing epidemic-style protocols usually achieve 100\% delivery, but they  waste resources by sending a large number of messages on the network,
whereas our approach succeeds to send only the amount of messages necessary to inform the
desired percentage of hosts in the given time. 


\section{Dissemination Techniques based on Epidemic Models}
\label{design}
In this section we introduce the mathematical models at the basis of
the design of the communication API presented in
Section~\ref{middleware}.  In order to model the message replication
mechanisms, we exploit mathematical models that have been devised to
describe the dynamics of infections in human
populations~\cite{AndMay92}. The study of mathematical models of
biological phenomena has been pioneered by Kermack and McKendrick in
the first half of the last century. In the last years,   
researchers have focused their interest in the  
modeling of disease spreading in networks characterised by
well-defined structural properties ~\cite{BBPV05,boccaletti06}.

\begin{figure}[t]
 \begin{minipage}[th]{0.45\textwidth}
   \centering
    \includegraphics[width=\textwidth]{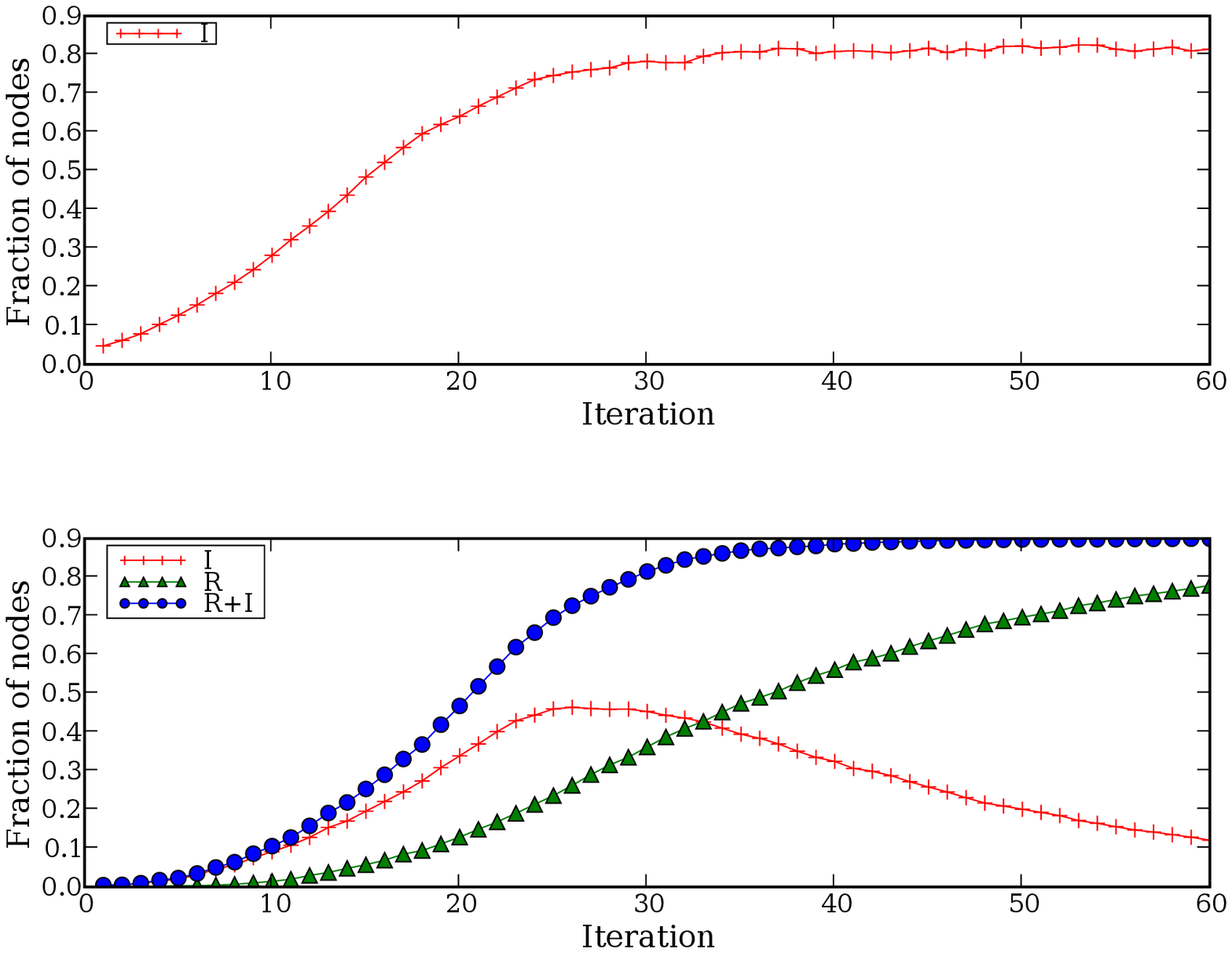}
    \caption{Infection spreading in the SIS (top) and SIR (bottom) model,  
under equivalent conditions. The number of infected (infected, removed and 
infected + removed) individuals is reported as a function of time for 
the SIS (SIR) model (with $\gamma=0.05$ and tuning the value of $\beta$ in order to have 
an infection of 100\%).}
    \label{comparation_infection}
 \end{minipage}
 \ \hspace{2mm} \hspace{3mm} \
 \begin{minipage}[th]{0.45\textwidth}
   \centering
    \includegraphics[width=\textwidth]{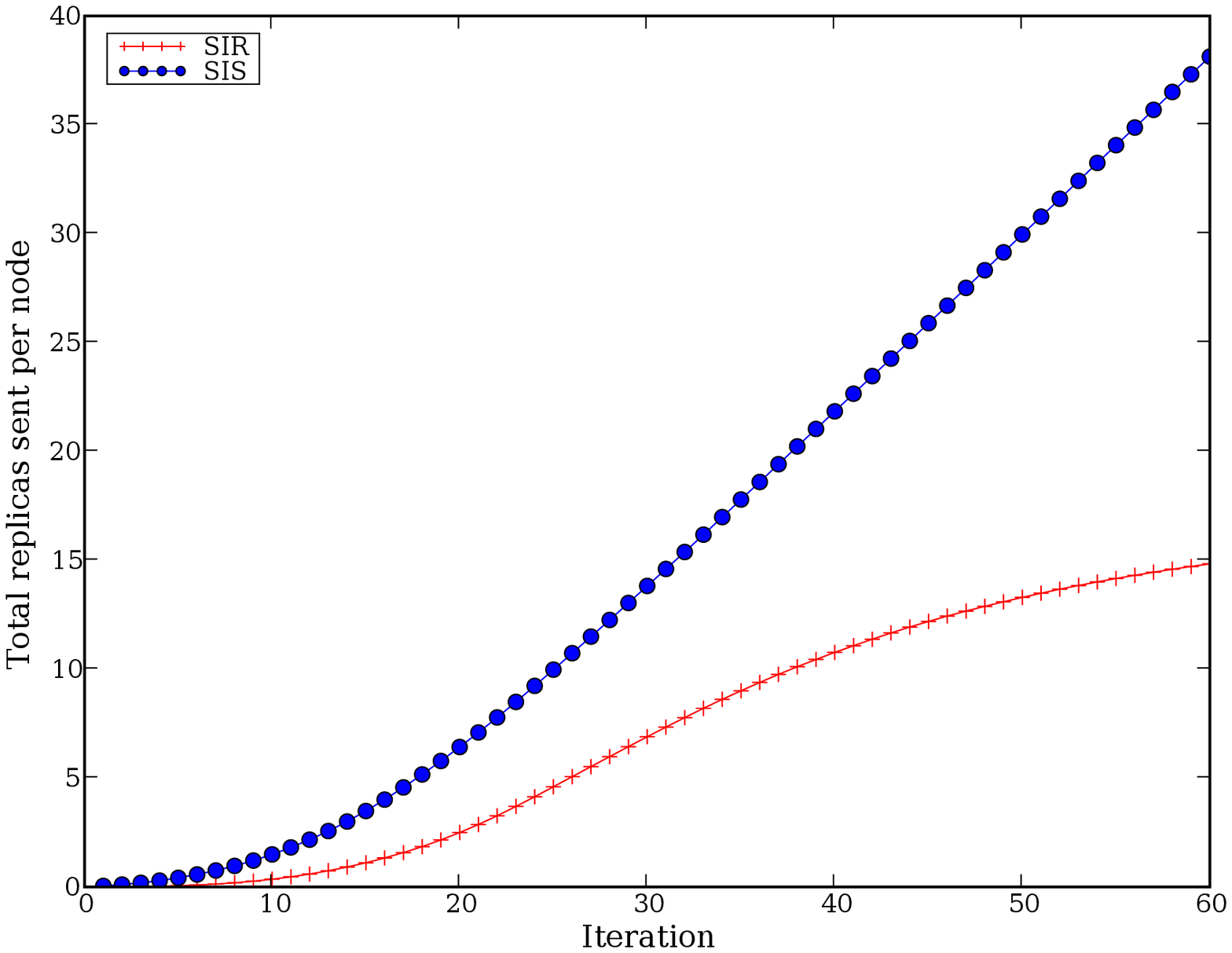}
    \caption{Number of replicas per host per message  
for the SIS and the SIR model. Same conditions as in Fig.1} 
    \label{comparation_msg}
 \end{minipage}
\end{figure}

According to the classic Kermack and McKendrick model, an individual
can be in three states: \textit{infected}, (i.e., an individual is
infected with the disease)
\textit{susceptible} (i.e., an individual is prone to be infected) and
\textit{removed} (i.e., an individual is immune, as it recovered from the
disease).  This kind of model is usually referred to as the
Susceptible-Infected-Removed (SIR) model~\cite{AndMay92}.  Removing
the possibility of permanently recovering from the disease a
different version of the model is obtained, according to which
individuals can exist in only two possible states, \textit{infected}
and \textit{susceptible}. In the literature, this model is usually
referred to as Susceptible-Infected-Susceptible (SIS)
model~\cite{AndMay92}.
The SIR model can guarantee the same delivery of the SIS model with a
substantially lower number of messages as shown by the generic
epidemic process depicted in Figures \ref{comparation_infection} and
\ref{comparation_msg}. This is due to the fact that the model
introduces the possibility of having hosts that are recovered, i.e.,
hosts that will not participate in spreading the infection after
having receiving a message $M$ and deleted it from the buffer. In
other words, in the SIR model the number of broadcasting nodes
decreases after a given peak of infected nodes; instead in the SIS
model, the number of broadcasting nodes at the end of the infection is
(approximately) equal to the number of nodes to be infected (i.e.,
desired percentage of nodes in the \verb+epcast+ primitive).

In the remainder of this paper we will substitute the term \textit{individual},
used by epidemiologists, with the term \textit{host}. A host is considered
infected if it holds the message and susceptible if it does not. If the message
is deleted from buffer of the host, the host becomes recovered and cannot be infected by the
same message anymore.  The information is spreaded among all infectives and
recovered, while susceptibles are still unaware of that: it is now clear that
the dissemination results depend on both infectives and recovered hosts, since
these are the actual recipients of the messages that have been sent. It is useful to define a host as
\textit{reached} if it is either an infective or a recovered, since in both
cases it has already received the message.  Moreover, it is worth noting that
only infectives contribute to message replication and spreading, while recovered
hosts do not. 

The main assumptions of our model are the following:
\begin{itemize}
\item all susceptibles in the population are equally at risk of infection from
any infected host (this hypothesis is usually defined by epidemiologists as
\textit{homogeneous mixing});
\item all infectives in the population have equal chances to recover;
\item the infectivity of a single host, per message, is constant\footnote{Note that the infectivity  per single message (i.e., a disease) is constant, but not per single host. In other words, a host usually stores messages characterised by different infectivities in its buffer.};
\item the initial number of the nodes in the network is known \textit{a priori} by each host\footnote{The initial number of hosts can be usually
estimated in occasion of sport events, rallies, etc. for example by evaluating
the seating capacity of the venues or the size of the area when the event takes
place. Statistical data are also usually available for many application
scenarios, such as number of passengers that uses a station or an airport in a
certain time of the day, etc. Alternatively, this number can be estimated using
distributed algorithms for the calculation of the approximated network size such
as~\cite{JM04}.};
\item every host collaborates to the delivery process and no malicious nodes are present;
\item each node has a buffer of the same size;
\item the number of hosts is considered constant during the spreading of the
infection\footnote{This is a realistic assumption, since users usually require
that the information will be disseminated in a limited time.}; 
\end{itemize}

Under the assumptions above, the system dynamics, in the case of a scenario
composed of $N$ active hosts, can be approximately\footnote{This is rigorously
justifiable in a network only 
for complete graphs in large population limit. However, the model provides a
good approximation also in scenarios composed of a limited number of hosts.}
described by the following system of non-linear differential equations~\cite{AndMay92}:
\begin{equation} 
\label{SIRmodel} 
\left \{ 
\begin{array}{l}

\frac{dS(t)}{dt}=-\beta S(t) I(t) \\

\frac{dI(t)}{dt}= \beta S(t) I(t) - \gamma I(t)\\

\frac{dR(t)}{dt}= \gamma I(t) \\

S(t)+I(t)+R(t)=N \\ 
\end{array} 
\right.  
\end{equation}
where $S(t), I(t), R(t)$ are respectively the number of susceptible,
infectives and removed hosts at time $t$, $\beta$ is the average
number of contacts with susceptible hosts that leads to a new infected
host per unit of time per infective, and $\gamma$ is the average rate
of removal of infectives per unit of time per infectives in the
population.  The equations of the system state that the decaying rate
of susceptibles and the growth rate of infectives are affected only by
the infectivity $\beta$, the number of susceptibles $S(t)$ and the
number of infectives $I(t)$; the decaying rate of infectives and the
relative growth of recovered is proportional to the removal rate
$\gamma$ and the number of infectives $I(t)$.  The last equation
states that actually only two equations are needed to completely define
the problem, since the sum of the number of hosts in the 
three classes is constant. 
We furthermore set the initial conditions: $S(0)= N-1$, $I(0)=1$, 
and $R(0)=0$, with the condition $I(0)=1$ representing
the first copy of the message that is inserted in its buffer by the
sender.

The numerical solution of system (\ref{SIRmodel}) can be easily
obtained by standard ODE solver routines. This allows to compute the
number of infectives and recovered at instant $t$ as a function of the
infectivity $\beta$ and of the removal rate $\gamma$. The value of
$\gamma$ is usually fixed by the local properties of the
hosts~\footnote{If overflow phenomena do not occur (i.e., in the case
of sufficiently large buffers), the model can be simplified with
$\gamma=0$ and, therefore, no host will never become
recovered.}. Instead, the value of $\beta$, that is the fundamental
parameter of the message replication algorithm, can be tuned in order
to have, after a specific length of time $t^*$, a number of reached
hosts (i.e., hosts that have received the message) equal to $I(t^*) +
R(t^*)$ or, in other words, a fraction of reached hosts equal to
$(I(t^*) + R(t^*))/N$.

In order to effectively exploit the model just described, the actual
connectivity of each host has to be taken into account.  We will
assume a mobile system with a {\em homogeneous network structure}, described
by a connectivity distribution $P(k)$, strongly peaked at an average
value $\langle k \rangle$.  This is a realistic assumption in cases
characterized by a high density of hosts, and where the movement is
well described as an uncorrelated random process, such as in large
outdoor spaces (i.e., squares, stations, airports or around sport
venues)~\cite{GKSG03,MM06}. In this case, the degree $k$ of each node
can be approximated quite precisely with the average degree $\langle k
\rangle$.  In order to include the effect of the connectivity on the
spreading, the system~(\ref{SIRmodel}) can be rewritten by
substituting $\beta$ with $\lambda \dfrac{\langle k
\rangle}{N}$~\cite{BBPV05}:
\begin{equation} 
\label{enhancedSIRmodel} 
\left \{ 
\begin{array}{l}

\frac{dS(t)}{dt}=-\lambda \dfrac{\langle k \rangle}{N} S(t) I(t) \\

\frac{dI(t)}{dt}= \lambda \dfrac{\langle k \rangle}{N} S(t) I(t) - \gamma I(t)\\

\frac{dR(t)}{dt}= \gamma I(t) \\

S(t)+I(t)+R(t)=N \\ 
\end{array} 
\right.  
\end{equation}
where $\lambda$ represents the probability of infecting a neighbouring
host during a unit of time, and $\frac {\langle k \rangle}{N}$ gives the
probability of being in contact with a certain host. In other words,
in this model, by substituting $\beta$ with $\lambda
\frac {\langle k \rangle}{N}$, we have separated, in a sense, the event of being
connected to a certain host and the infective process~\cite{BBPV05}.

In conclusion, the main idea is to calculate the value of $\lambda$ as
a function of $I(t^*)+R(t^*)$ and $\langle k \rangle$.  It is also 
interesting to note that in homogeneous networks, every host knows its
value of $k$ and, consequently, it has a good estimate of $\langle k
\rangle$. We will exploit this property to tune the spreading of
message replicas in the system.

\section{Implementation}
\label{implementation}

Every time the middleware primitive defined in Section~\ref{middleware}
is invoked, the middleware calculates the value of the infectivity $\lambda$
that is necessary and sufficient to spread the information to the desired
fraction of hosts in the specified time interval (specified in the field \verb+percentageOfHosts+ of the \verb+epcast+ primitive), by evaluating the current average
degree of connectivity and the current removal rate of messages from the buffer.
The message identifiers, the value of the calculated infectivity, the timestamp
containing the value specified in \verb+time+ expressing its temporal validity
are inserted in the corresponding headers of the message in the
\textit{infectivity} field. Then, the message is inserted in the local buffer. 

The epidemic spreading protocol is executed periodically with a period equal to
$\tau$. With respect to the calculation of the message infectivity, we assume $\tau$ as time unit in the formulae presented in
Section~\ref{design}. In other words, assuming, for example, $\tau=10$, a
timestamp equal to one minute corresponds to six time units. The value of $\tau$
can be set by the application developer during the deployment of the platform.
Clearly, the choice of the values of $\tau$ influences the accuracy of the
model, since it relies on a probabilistic process. For this reason, given a
minimum value of timestamp equal to $t_{MIN}$, developers should ensure $\tau <<
t_{MIN}$.  The number of rounds will be equal to $t^* / \tau$. For the Law of
the Large Numbers, we obtain a better accuracy of the estimation of the evolution
of the epidemics as the number of rounds (i.e., from a probabilistic point of
view, the number of trials) increases.

Every $\tau$ seconds each infected host broadcasts the message and its
neighbours receive the message. If the message is not already present in their
buffer, they store it with a probability $\lambda$: moreover, they will not store
it if the message has been already present in buffer in the past, although it is
not present at current time. This behaviour maps quite well the SIR epidemic spreading
model, since a node receives a new message, actively spreads it for some time
and then it deletes the message from the buffer (i.e. to make room for new
messages), never accepting it again. Therefore, a node has to store the identifiers of all messages received in a defined time window, which
is a reasonable given the limited occupation of the vector of the message identifiers.

\section{Evaluation}
\label{evaluation}
\subsection{Analytical Evaluation}
\label{analytical}
\label{estimationnumberofmessages}
An interesting quantitative parameter is the total number of messages
needed to disseminate messages to a certain percentage of hosts. A message is
broadcasted by an infective host in every round: as soon as the host deletes the
message it does not accept the same message again.
 
Considering an infection process repeated for a number of times equal to $r$
number of rounds, indicating with $t_{r}$ the time length of the $r^{th}$ round,
the total number of replicas per single type of message can be estimated as
follows:
\begin{equation}
 \text{ Number Of Replicas} = \int_{t=0}^{t=t_{r}} I(t) dt
\label{numberofreplicasequation}
\end{equation}

From a graphical point of view, the number of copies is equal to the area under the curves in Figure \ref{comparation_infection} and \ref{comparation_msg}. A comparison between SIR-
and SIS-based protocols shows that while for both cases the formula
\ref{numberofreplicasequation} helds, in the former case the total number of
replicas sent is much lower. This is the result of the recovering process, which
enables hosts to stop message spreading when the epidemics is already growing
but, at the same time, still guarantees that the final result will be guaranteed.

\subsection{Experimental Evaluation}
\label{experimental}
\subsubsection{Description of the Simulation}
In order to test the performance of these techniques, we defined a square simulation area with a side of 1 km
and a transmission range equal to 200 m. The simulation was set to run several
runs for each mobile scenario in order to obtain a statistically
meaningful set of results (with a maximum 5\% error). All simulations are
written in Python using NetworkX \footnote{
\tt{http://networkx.lanl.gov}}, a package for the
creation, manipulation, and study of the structure, dynamics, and functions of
complex networks.
We analysed scenarios characterised by different number of hosts (more precisely
64, 128, 256, 512). These input parameters model typical deployment settings of
mobile ad hoc networked systems.  We do not model explicitly the failures in the
system, since we assume that during the infection process, the number of hosts
remains constant. 

The movements of the hosts are generated using a Random Way-Point
mobility model~\cite{CBD02}; every host moves at a speed that is
randomly generated by using a uniform distribution. The range of the
possible speeds is $[1,6]m/s$. We selected this mobility model, since
as discussed in~\cite{GKSG03}, its emergent topology has a Poisson
degree distribution.  Therefore, in this scenario, the properties of
the network can be studied with a good approximation by assuming a
homogeneous network model.  The accuracy of the approximation
increases as the density of population increases, since, considering
the finite and limited simulated time, we obtain a scenario
characterised by a time series of degree of connectivity values with lower variance. Moreover, the so-called border
effects, due to the host that moves at the boundaries of the simulated
scenarios, have less influence as the density of population increases.

Each node uses a buffer of 5 messages, managed as a FIFO (first in first out)
queue, and 20
different messages are sent in the initial round by random chosen nodes.


\begin{figure}[t]
\centering
 \begin{minipage}[t]{0.45\textwidth}
   \centering
    \includegraphics[width=1\textwidth]{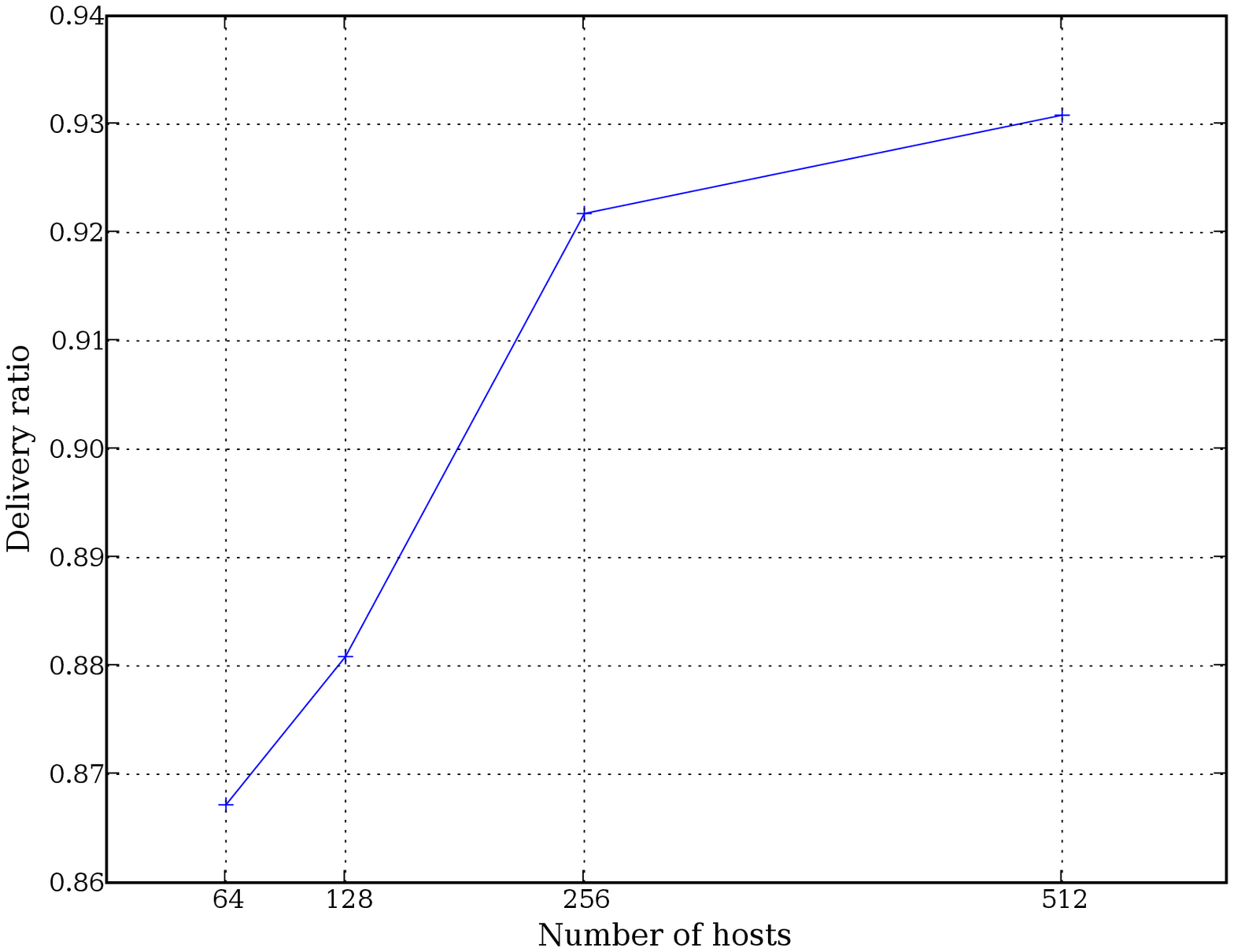}
    \caption{Delivery ratio vs population density with desired reliability equal to 100.}
    \label{deliveryratiovspopdensity_a}
 \end{minipage}
 \ \hspace{2mm} \hspace{3mm} \
 \begin{minipage}[t]{0.45\textwidth}
   \centering
    \includegraphics[width=\textwidth]{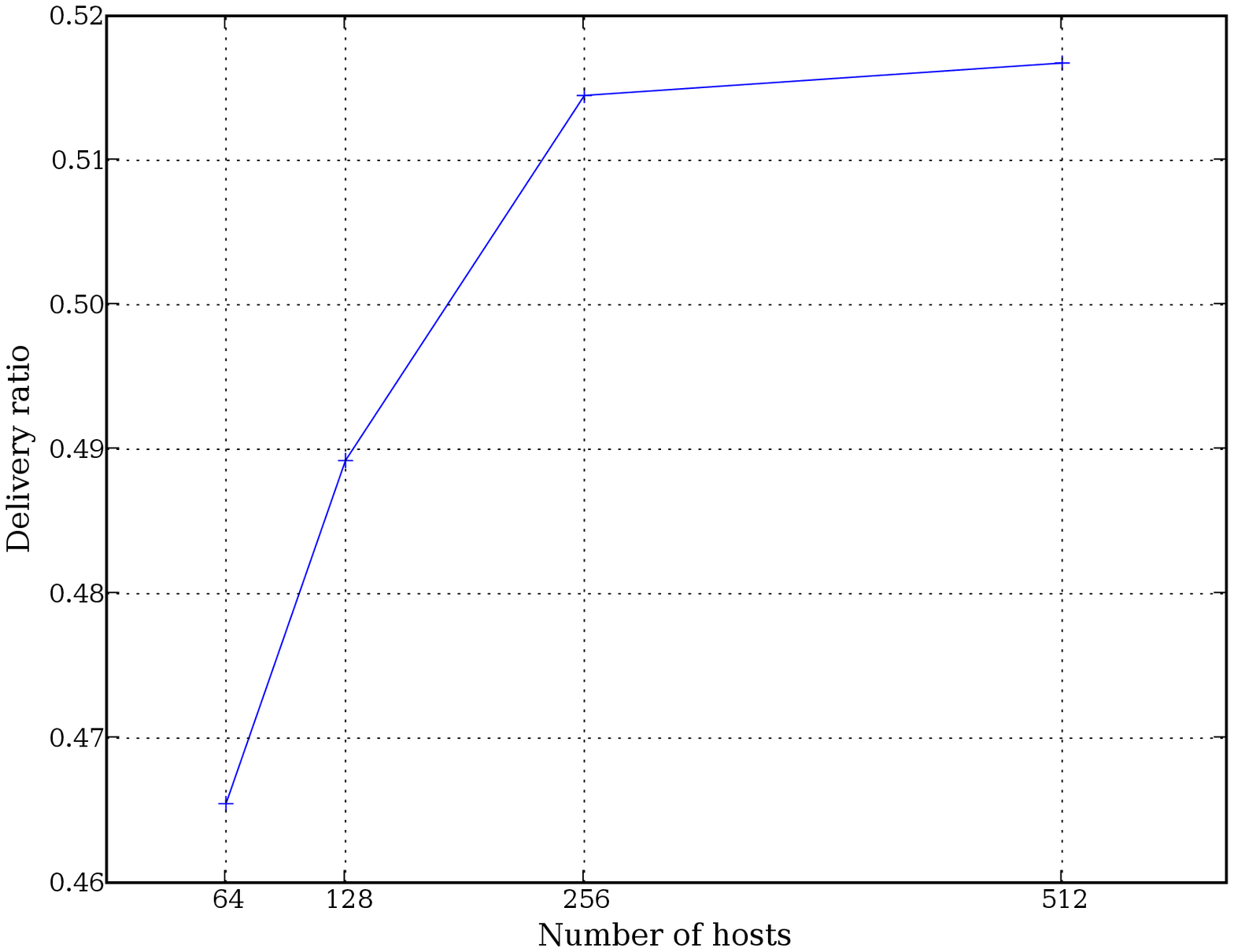}
    \caption{Delivery ratio vs population density with desired reliability equal to 50.}
    \label{deliveryratiovspopdensity_b}
 \end{minipage}
\end{figure}

%
%
\begin{figure}[t]
\centering
 \begin{minipage}[t]{0.45\textwidth}
   \centering
    \includegraphics[width=\textwidth]{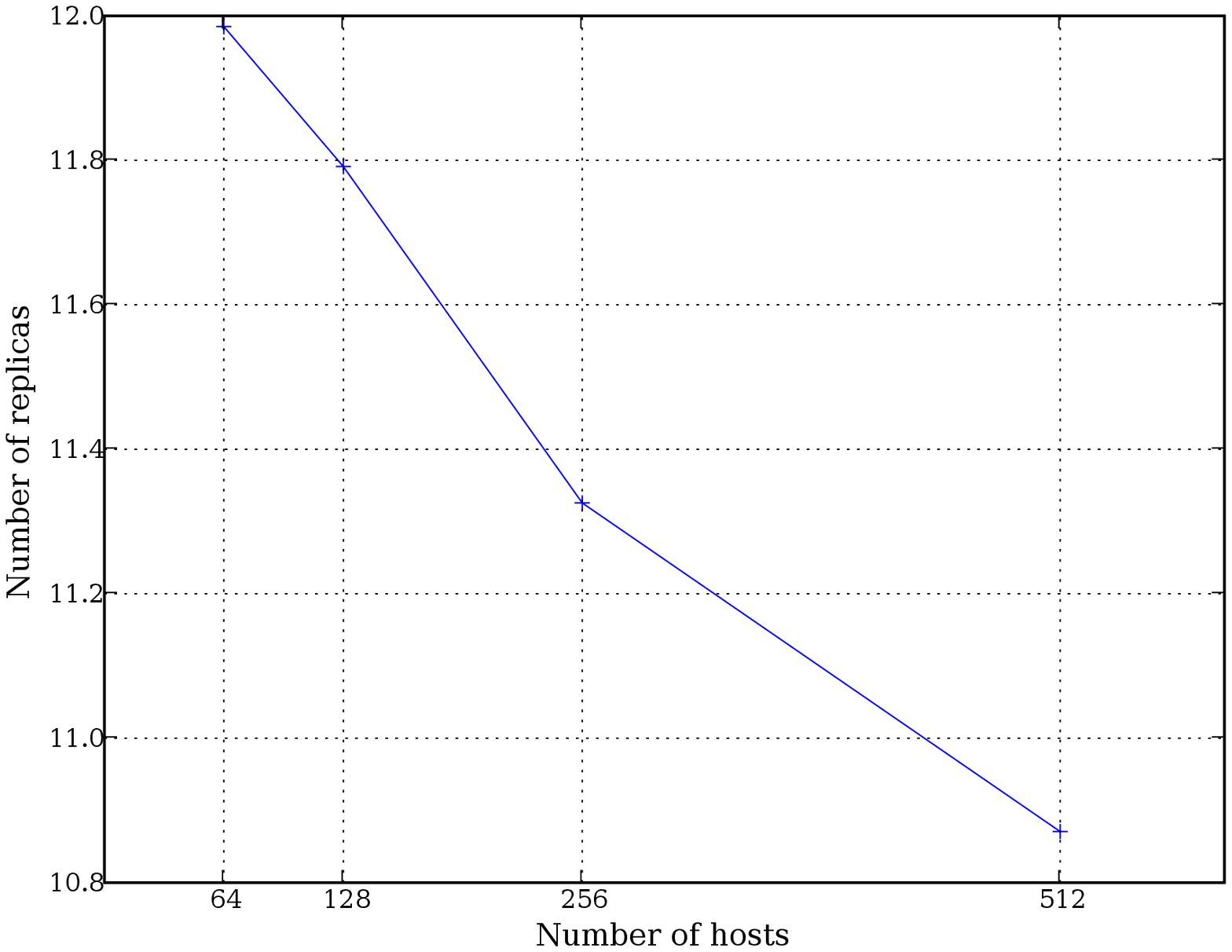}
    \caption{Number of replicas per host per message vs population density with desired reliability equal to 100.}
    \label{numberofreplicasvspopdensity_a}
 \end{minipage}
 \ \hspace{2mm} \hspace{3mm} \
 \begin{minipage}[t]{0.45\textwidth}
   \centering
    \includegraphics[width=\textwidth]{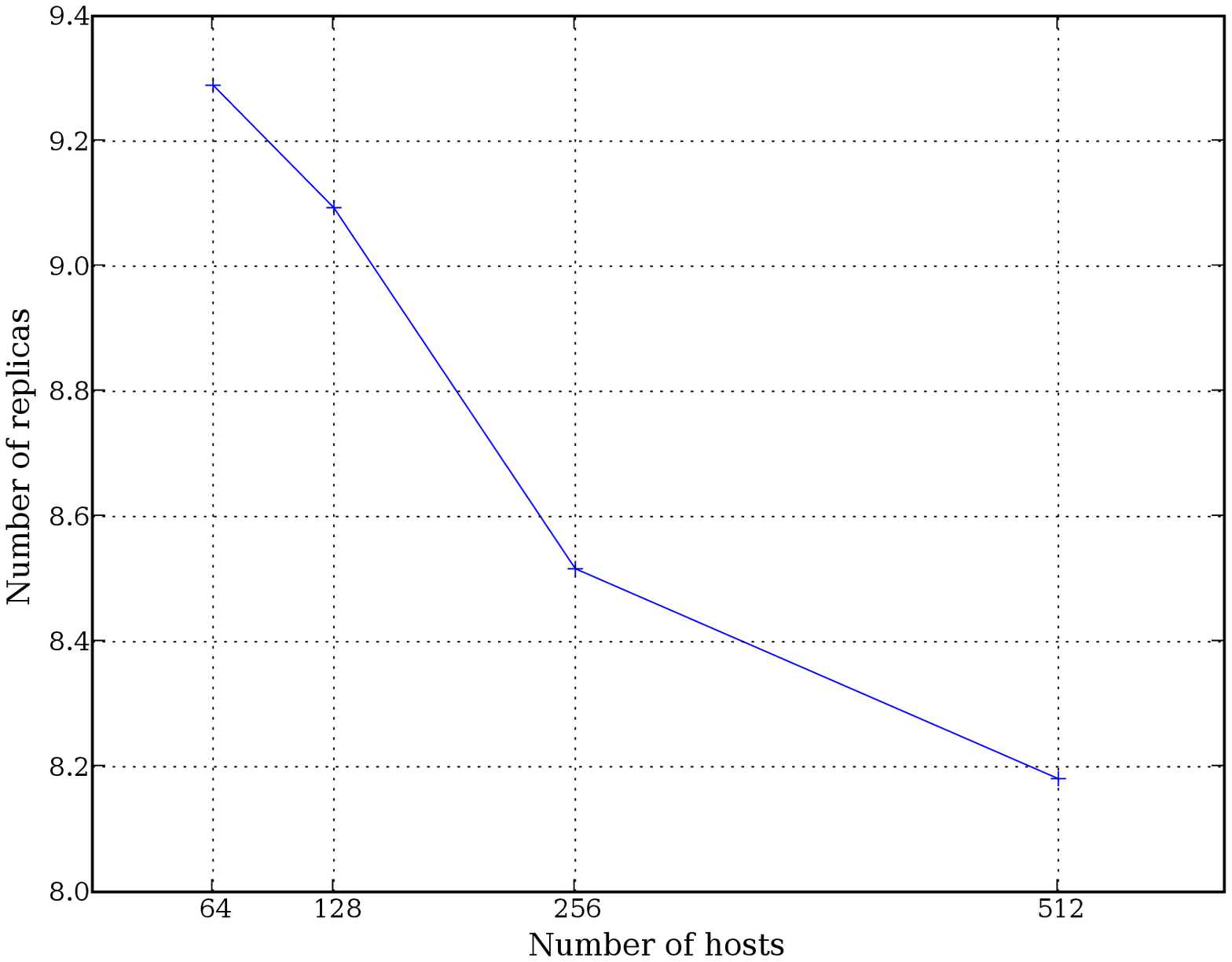}
    \caption{Number of replicas per host per message vs population density with desired reliability equal to 50.}
    \label{numberofreplicasvspopdensity_b}
 \end{minipage}
\end{figure}

\subsubsection{Analysis of Simulation Results}
In this subsection we will analyse the results of our simulations,
discussing the performance of the proposed techniques. We will study
the variations of some performance indicators, such as the delivery
ratio and the number of messages sent as functions of the density of
hosts (i.e., the number of the hosts in the simulation area).
 
Figure~\ref{deliveryratiovspopdensity_a} and~\ref{deliveryratiovspopdensity_b}
show the delivery ratio (i.e., the desired percentage of hosts in the \verb+epcast+ primitive) in terms of population density, for the case of a
desired percentage of hosts equal to 100 and 50, respectively, with $t^*=10min$. 
The performance in terms of delivery ratio are close to the desired ones. Also in this case, the better approximation of the assumption of
homogeneous network, obtained when the density of population increases, leads to
better results (i.e., a more accurate estimation) for the case of 512 nodes.

%

The number of replicas per host per message are plotted in
Figure~\ref{numberofreplicasvspopdensity_a}
and~\ref{numberofreplicasvspopdensity_b}.  These diagrams
illustrate the scalability of our approach, since the number of replicas is
slightly decreasing when more nodes are added.

\subsubsection{Evaluation with Dartmouth traces}
In order to evalute our approach on real data we run simulations using a source
of data describing how real users move between different locations, i.e.
wireless access points. A large amount of traces from the
802.11b campus network of Dartmouth College is available through the CRAWDAD project
\cite{dartmouth}.
We selected all the contacts between 9 am and 6 pm in one day during the term (Wednesday 21 April 2004), discarding
contacts with duration less than 60 seconds. Two users are connected only if
they are associated with the same access point during a time slot: epidemics
spreading is therefore performed among users co-located with access points.
Our resulting data set had 2201 unique MACs and 11572 contacts with all access points. We assume that each MAC address corresponds to a unique user.
The other simulation parameters are the same of the previous analysis. 
\begin{table}[t]
   \centering
\begin{tabular}{|c|c|c|c|}
\hline
Type & Desired fraction & Delivered fraction & Messages sent \\
\hline
\hline
epcast & 0.50 & 0.43 & 17132 \\ 
epcast & 0.75 & 0.68 & 24738 \\ 
epcast & 1.00 & 0.90 & 32475 \\ 
epcast(heterogeneous) & 1.00 & 0.90 & 57342 \\ 
Epidemic ($\beta = 0.25$) & 1.00 & 0.64 & 95969 \\ 
Epidemic ($\beta = 0.50$) & 1.00 & 0.87 & 121873 \\ 
Epidemic ($\beta = 1.00$) & 1.00 & 0.92 & 155446 \\ 
\hline
\end{tabular}
\caption{Comparation of performances on the real dataset of Dartmouth College traces}
\label{tab_results}
\end{table}
In Table \ref{tab_results} we show the performances of our approach: the
percentage of host actually reached is slight less than the desired
fraction of population and this can be explained by observing that these hosts are not
always connected during all the simulation time and may be easily
absent from the underlying network.
In other words, the underpinning hypothesis of  the epidemic spreading model that we are using are only approximately satisfied.
We run a simulation with a standard epidemic approach where infectivity is not
tuned using the SIR model but it is set to 0.25, 0.50 and 1.00 respectively. It is interesting to note that the number of messages is in all three cases higher; only the case with infectivity equal to 1.00, the standard epidemic protocol is able to reach all the hosts. This is also demonstrate how it is difficult to choose the right value of the infectivity in a purely heuristic way to reach all the hosts of the system.

We run also some simulations using a dataset from the National University of Singapore\cite{singapore}, which contains contact pattern of 22341 students
inferred from the information on class schedules and class rosters for the
Spring semester of 2006. Two students are connected if they attend the same
class during a time slot. However, in this dataset a large fraction of students is not included in the
instantaneous underlying network, since they are not attending any class.
The result is that in this case the epidemics fails to start using our model based on the assumption of homogeneous mixing.
Additional virtual point of aggregation can be included in the simulations,
grouping a percentage of the students that are not attending lectures during a
particular timeslot: this modification ensures homogeneous mixing, providing good results for our algorithm. However, this is only a conjecture given the
nature of the traces.

\subsubsection{Heterogeneous Networks}
The results and the solutions discussed in this paper rely on the assumption of
homogeneous networks, that are emerging from the random movements of the nodes.
We now show that the proposed approach can be  extended to the
general case of {\em heterogeneous networks} ~\cite{BBPV05,boccaletti06}. 
These structures are emerging in
presence of small clusters of people or communities. 

For heterogeneous networks the approximation $k \approx \langle k \rangle$ is
not valid. However, the same probabilistic communication primitives introduced
in Section~\ref{middleware} could be used, with a different semantics.  This
relies on the following observations: given $k$ fluctuating in the range
$[k_{MIN},k_{MAX}]$, we observe that for a value of the infectivity
corresponding to $k=k_{MIN}$, the obtained spreading of the infection
$I(t^*,k_{MIN})$ will always be greater than the one obtained with another $k$.
In other words, if $k_{MIN}$ is selected in the calculation of the value of the
infectivity, the value of \verb+Reliability+ can be considered approximately as
a guaranteed lower bound of the reliability level. 

The value of $k_{MIN}$ can be
dynamically retrieved and set by the middleware by monitoring the connectivity
of the hosts composing the mobile system. We plan to investigate these adaptive mechanisms
further in the future.


\section{Concluding Remarks}
\label{concluding}
In this paper we have shown how models of epidemic spreading in
complex networks can be applied effectively to the problem of
disseminating information to subset of hosts (or to all the hosts) in
a wireless network, controlling at the same time the number of the
copies in the system. We have presented an analytical and experimental
evaluation of our approach using a synthetic random model and real
traces, showing the effectiveness of our approach. 

\medskip
{\bf Acknowledgements} Cecilia Mascolo and Mirco Musolesi acknowledge the support of EPSRC through the CREAM Project. Salvo Scellato thanks UCL for the financial support as Visiting Student. 

{
\tiny
\bibliographystyle{abbrv}
\bibliography{bioinspired}
}
\end{document}